\documentclass{elsart}
\usepackage{graphics}
\usepackage{graphicx}
\usepackage{amssymb}
\begin{document}

\begin{frontmatter}

\title{Interplay between $3d$ and $4f$ magnetism in CeCoPO}

\author[MPI]{C. Krellner\corauthref{CK}},
\ead{krellner@cpfs.mpg.de}
\author[MPI]{U. Burkhardt}, and
\author[MPI]{C. Geibel}
\address[MPI]{Max Planck Institute for Chemical Physics of Solids, D-01187 Dresden, Germany}

\corauth[CK]{Corresponding author. Tel/Fax: +49 351 46462249}

\begin{abstract}
The ground state properties of CeCoPO, a homologue of the new high temperature superconductors
$Ln$Fe$Pn$O$_{1-x}$F$_x$, were studied by means of susceptibility, specific heat, and resistivity measurements on polycrystals. The observation of a well defined Curie-Weiss behavior above 230\,K with $\mu_{\rm eff}=2.9\,\mu_B$ and a ferromagnetic ordering below  $T_C=75$\,K is similar to what was observed in LaCoPO and points to magnetism of the Co-$3d$ electrons. However, the Ce-ions are on the border to magnetism with a Kondo scale of $T_K\sim 40$\,K and show an enhanced Sommerfeld-coefficient of $\gamma\sim200$\,mJ/molK$^2$.
\end{abstract}

\end{frontmatter}

\section{Introduction}
\label{Intro}
The compound series \textit{LnTPn}O ($Ln$: lanthanides, $T$: transition metal, $Pn$: P or As) have started to attract considerable attention because of the recent discovery of superconductivity with a transition temperature $T_c$ exceeding 50\,K in the $Ln$FeAsO$_{1-x}$F$_x$ series of compounds \cite{Kamihara:2008},\cite{Chen:2008},\cite{ChenNature:2008}, being the highest $T_c$ except for cuprate systems. While the recent reports focus on the importance of electronic correlation effects due to $3d$-electrons close to a magnetic state, the homologous compounds with $Ln$ = Ce are attractive candidates for strong correlation effects induced by the $4f$-electrons. Thus, last year we presented a detailed study of the properties of CeRuPO and CeOsPO \cite{Krellner:2007}, \cite{Krellner:2007b}, \cite{Krellner:2008} and demonstrated that the former one is a rare example for a ferromagnetic (FM) Kondo lattice system with a FM ordering temperature $T_C =15$\,K and a Kondo temperature $T_K\sim10$\,K, while the latter one shows antiferromagnetic order of stable trivalent Ce-ions below $T_N=4.5$\,K. More recently, we have shown that CeFePO is a heavy Fermion metal close to a FM instability, dominated by the magnetism of the $4f$-electrons \cite{Bruning:2008}. Therefore, it was natural to look also for the physical properties of the related compound with $T$ = Co.
 
In this report, we present a first study of the basic physical properties of CeCoPO using susceptibility $\chi(T)$, specific heat $C(T)$, and resistivity $\rho(T)$ measurements as well as x-ray absorption near-edge spectroscopy structure (XANES) at the Ce $L_{\rm III}$ edge. The results reveal a FM ordering of the Co-ions at $T_C=75$\,K in analogy to the FM order at 43\,K in LaCoPO \cite{Yanagi:2008}. An enhanced Sommerfeld-coefficient of $\gamma\sim200$\,mJ/molK$^2$ and the temperature dependence of the resistivity reveal a finite Kondo scale of the $4f$-moments of order $T_K\sim40$\,K. In addition, we confirm the reported data of the magnetic properties of polycrystalline LaCoPO. 

\section{Experiment}
\label{Ex}
Polycrystalline samples of CeCoPO and LaCoPO were synthesized using a Sn-flux method in evacuated quartz tubes as described elsewhere \cite{Krellner:2007}, \cite{Kanatzidis:2005}. Several powder x-ray diffraction patterns recorded on a Stoe diffractometer in transmission mode using monochromated Cu-K$_{\alpha 1}$ radiation ($\lambda= 1.5406$\,\AA) confirmed the formation of $Ln$CoPO with ZrCuSiAs structure type (space group: $P4/nmm$; $Z=2$). The lattice parameters refined by simple least square fitting  for CeCoPO $a = 3.924(3)$\,\AA, $c = 8.223(5)$\,\AA\, and LaCoPO $a = 3.968(3)$\,\AA, $c = 8.379(5)$\,\AA\, were found to be in good agreement with the reported structure data \cite{Yanagi:2008}, \cite{Zimmer:1995}. In addition, energy dispersive x-ray spectra of CeCoPO and LaCoPO performed on a scanning electron microscope (Philips XL30) with Si(Li)-x-ray detector show similar strong intensities of the oxygen line which confirm a substantial oxygen content of both phases. However, they revealed diamagnetic CoP$_2$ \cite{Jeitschko:1984} as impurity phase and some oxide phases which are present only with small concentration, so that they could not be observed in the x-ray diffraction patterns. $\chi(T)$ measurements were performed in the temperature range $2-350$\,K in a commercial Quantum Design (QD) magnetic property measurement system (MPMS), the $T$-dependent measurements were performed after zero-field cooling the sample. The resistivity was determined down to 0.4\,K using a standard AC four-probe geometry in a QD physical property measurement system (PPMS). The PPMS was also used to measure $C(T)$ with a standard heat-pulse relaxation technique. XANES-spectra near the Ce $L_{\rm III}$ edge were taken at 300\,K in transmission geometry at the EXAFS beamline A1 of HASYLAB at DESY, Germany. The wavelength selection was realized by means of Si(111) double crystal monochromator, which allowed an experimental resolution of approximately 2\,eV (FWHM) at the Ce $L_{\rm III}$ threshold of 5723\,eV.

\begin{figure}[t]
\begin{center}
\includegraphics[width=10cm]{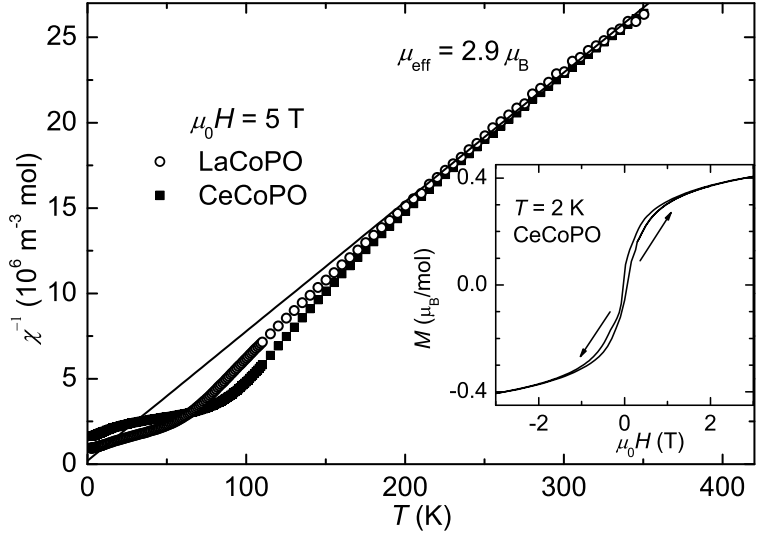}
 \caption{Inverse susceptibility as function of temperature for CeCoPO (squares) and LaCoPO (circles) measured in a magnetic field of $\mu_0H=5$\,T. Both compounds show similar Curie-Weiss behavior above 230\,K with $\mu_{\rm eff}=2.9\,\mu_B$ and $\Theta_W=-3$\,K (straight line). Inset: The magnetization of CeCoPO presents a small hysteresis at small magnetic fields and a nearly saturated moment at higher fields.}
\label{FigChi1}
\end{center}
\end{figure}

In Fig.~\ref{FigChi1} the inverse susceptibility is shown for both CeCoPO and LaCoPO. Above 230\,K both curves present similar Curie-Weiss behavior with an effective moment of $\mu_{\rm eff}=2.9\,\mu_B$ and a Weiss temperature of $\Theta_W=-3$\,K. Our results for LaCoPO are very similar to those of Ref.~\cite{Yanagi:2008} but important to analyze the magnetism in CeCoPO. The identical behavior of CeCoPO and LaCoPO in $\chi^{-1}(T)$ evidences that in both compounds the main contribution to the magnetism results from the $3d$-electrons of Co. However, in CeCoPO the Ce-ions are not completely unmagnetic as will be discussed later. In the inset of Fig.~\ref{FigChi1}, the magnetization $M(H)$ of CeCoPO is shown. $M(H)$ presents a small hysteresis at low field and a saturated moment of $\mu_{\rm sat}\sim0.4\,\mu_B$ at $\mu_0H\sim3$\,T, similar to what was observed in LaCoPO \cite{Yanagi:2008}. At higher fields, $M(H)$ increases further slightly which might be due to the paramagnetism of the Ce-ions.

\begin{figure}[t]
\begin{center}
\includegraphics[width=10cm]{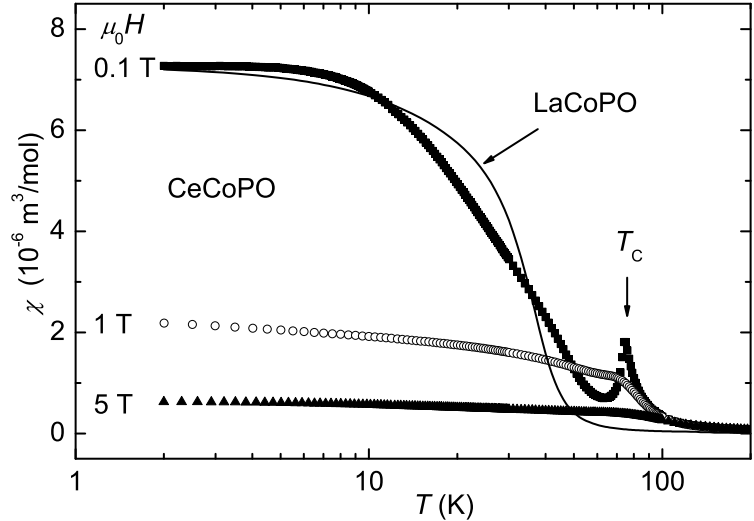}
 \caption{$\chi(T)$ of CeCoPO at various magnetic fields. The magnetic transition into a FM ordered state is visible at $T_C=75$\,K. For comparison, $\chi(T)$ for LaCoPO at $\mu_0H=0.1$\,T is presented as well (straight line).}
\label{FigChi}
\end{center}
\end{figure}

In Fig.~\ref{FigChi}, we present $\chi(T)$ of CeCoPO as a function of temperature at three different magnetic fields. The transition into a FM ordered state is visible by a sharp increase at $T_{C}=75$\,K and a strong field dependence of $\chi$ at low $T$. However, the peak at $T_C$ observed in low fields ($\mu_0H\leq 0.1$\,T) does not correspond to the expectation for a simple ferromagnet. Presently, it is not settled whether this is the result of measuring powder samples of a system with a complex anisotropic behavior or if it points to a more intricate magnetic structure with polarization effects on the Ce-ions by the internal magnetic fields of the Co-magnetism. The later scenario is more likely, because the $T$-dependence of $\chi$ for LaCoPO (line in Fig.~\ref{FigChi}), where no $4f$-electrons are present, behaves as expected for a simple ferromagnet.

\begin{figure}[t]
\begin{center}
\includegraphics[width=10cm]{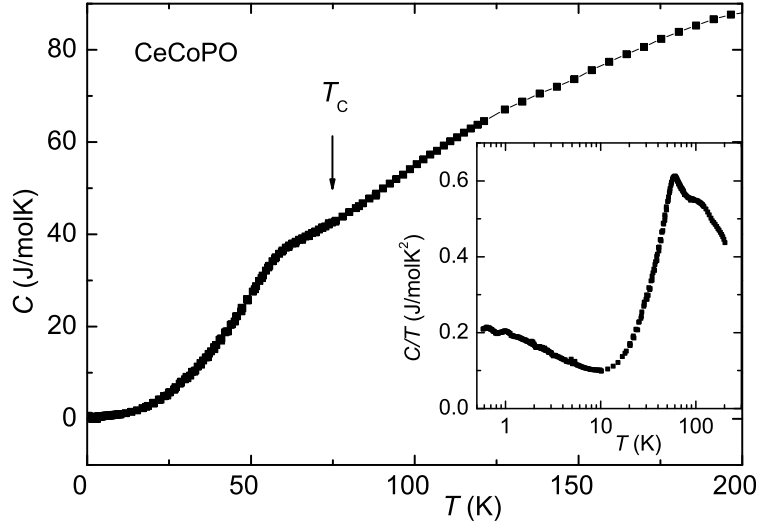}
 \caption{Temperature dependence of the specific
heat of CeCoPO. The phonon contribution is not subtracted in
this representation. A broad anomaly is visible below $T_C$ due to the FM order of CeCoPO. The inset shows the specific heat plotted as $C/T$ on a logarithmic temperature scale. Below $T=10$\,K, $C/T$ of CeCoPO increases logarithmically down to lower $T$, presenting an enhanced $\gamma\sim 200$\,mJ/molK$^2$ at 0.35\,K.}
\label{FigHC}
\end{center}
\end{figure}

The temperature dependence of the specific heat of CeCoPO is shown in Fig.~\ref{FigHC} without any subtraction of the phonons. Therefore, at high $T$, $C(T)$ converges to the classical limit of $4\cdot3R\sim 100$\,J/molK. An anomaly is visible on top of the phonon excitations just below $T_C$. This anomaly is rather broad, possibly due to inhomogeneities in our polycrystalline sample. In the inset of Fig.~\ref{FigHC} we have plotted the specific heat as $C/T$ on a logarithmic temperature scale. In this representation the peak due to the onset of FM order is more pronounced. Below $T=10$\,K, $C/T$ increases logarithmically and tends to saturate below $T=1$\,K at $\gamma\sim 200$\,mJ/molK$^2$. This value is strongly enhanced compared to the expectation for a normal FM Co-system. This enhancement can be attributed to $4f$-correlation effects resulting in the formation of heavy Fermions. The entropy at low $T$ was calculated by integrating $C/T$ over temperature. The entropy gain at 10\,K corresponds to about 25\% of $R\ln 2$, giving a rough estimation of the Kondo energy scale of order $T_K\sim 40$\,K. For the calculation, we have not subtracted any nonmagnetic contribution which for $T<10$\,K is negligible compared to the large value of $C/T$. As an example, the contribution of the nonmagnetic LaRuPO to the specific heat at $T = 1$\,K would only correspond to 2\% of the here measured specific heat \cite{Krellner:2007}.

\begin{figure}[t]
\begin{center}
\includegraphics[width=10cm]{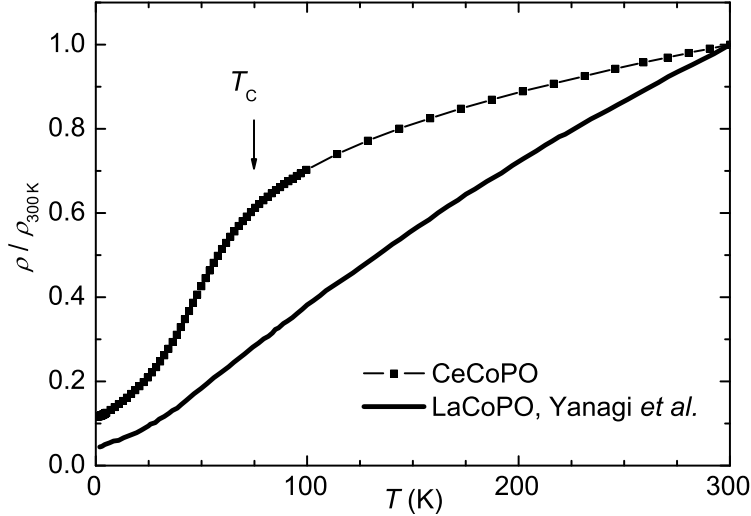}
 \caption{Temperature dependence of the resistivity of CeCoPO (squares) normalized to the room temperature value $\rho_{300\rm K}=5\,$m$\Omega$cm. For comparison, $\rho(T)$ of LaCoPO taken from Yanagi \textit{et al.} \cite{Yanagi:2008} is shown (straight line).}
\label{FigRho}
\end{center}
\end{figure}

In Fig.~\ref{FigRho} we present the resistivity data on a pressed powder-pellet of CeCoPO which was subsequently annealed for 150 hours at 700$^{\circ}$C, improving the conductivity of the sample by a factor of 20. However, the absolute value of $\rho_{300\,\rm K}=5\,$m$\Omega$cm is still rather high; we believe that this is not intrinsic of CeCoPO but due to the remaining granularity of the polycrystalline sample. To compare our results with $\rho(T)$ of LaCoPO from Yanagi \textit{et al.}, we have normalized our data to $\rho_{300\,\rm K}$. From Fig.~\ref{FigRho} it is obvious that $\rho(T)$ of CeCoPO differs significantly from the data of LaCoPO which decreases linearly with temperature with a small change in slope at the FM ordering temperature. In contrast, CeCoPO shows linear metallic behavior only above 100\,K, followed by a pronounced decrease which is a distinct feature of a Kondo lattice system. However, we can not exclude that this decrease is due to reduced spin-spin scattering because of the FM order at $T_C$, as both energy scales ($T_K\sim 40$\,K and $T_C=75$\,K) are comparable. Thermopower measurements on CeCoPO would be desirable to distinguish between both effects.

We have performed XANES measurements at the Ce$^{3+}$ edge to determine the valency of the Ce-ions to gain a deeper insight of the relevance of cerium to the magnetic behavior in CeCoPO. In the inset of Fig.~\ref{FigXAS} we have plotted the evolution of the tetragonal unit cell volume $V$ for the three $LnT$PO series ($T=$\,Co, Fe, Ru). The contraction of $V$ is typical for the $Ln$ series, and a smooth evolution is expected if Ce is in a trivalent state. Therefore, one would expect a stable, magnetic Ce$^{3+}$ in CeRuPO but a stronger tendency to intermediate-valent Ce-ions in CeFePO and CeCoPO \cite{Krellner}. This trend is confirmed in our XANES spectra shown in Fig.~\ref{FigXAS}. All three spectra are quite close to each other indicating, that the valency of Ce is not changing dramatically when going from Ru- to the Co-compound.  Above 5.73\,keV, the Co-compound show a slightly higher intensity compared to CeRuPO, but far away from the double-peak behavior in the formal pure Ce$^{4+}$-system CeO$_2$ \cite{Kaindl:1988}. From this we conclude a predominantly Ce$^{3+}$ for all three systems. Preliminary x-ray magnetic circular dichroism (XMCD) measurements on CeCoPO confirm this result showing that more than 90\% of the Ce-ions are in the $J=5/2$ configuration \cite{Goering:2008}.

\section{Discussion}\label{Dis}
Our results on CeCoPO evidence three interesting features which shall now be discussed: The interplay of the $3d$ and $4f$ magnetism, the comparison with the results obtained on CeRuPO and CeFePO, and the relation to the new high temperature superconductors $Ln$Fe$Pn$O$_{1-x}$F$_x$.

The presented results point to a FM ordering of the Co-ions in CeCoPO similar to the FM order in LaCoPO \cite{Yanagi:2008}; however, the unusual temperature dependence of $\chi(T)$ at $T_C$ together with an increasing magnetization for fields $B\geq 3$\,T are not expected for a simple FM and not observed in LaCoPO. On the other hand, the XANES spectra suggest the Ce-ions to be close to trivalent which is in agreement with the enhanced $\gamma$, and the temperature dependence of $\rho(T)$ which are typical for pronounced Kondo-interactions. Therefore, we suggest the following scenario for CeCoPO. The spins of the Co-ions ($3d$-electrons) order ferromagnetically at $T_C=75$\,K, the resulting internal fields partially polarize the Ce-ions ($4f$-electrons), which however remain also partially fluctuating down to lowest measured temperatures. One could even imagine that a Kondo screening increasing with decreasing temperature lead to a decreasing Ce-polarization with temperature. In order to get a deeper insight into the many-body phenomena in this compound, a more microscopic local probe should be used to study the magnetism. We therefore, started an investigation of CeCoPO with XMCD \cite{Goering:2008} as well as $^{31}$P NMR and $^{59}$Co NMR \cite{Bruning:2008a} to distinguish between $3d$ and $4f$ magnetism.

\begin{figure}[t]
\begin{center}
\includegraphics[width=10cm]{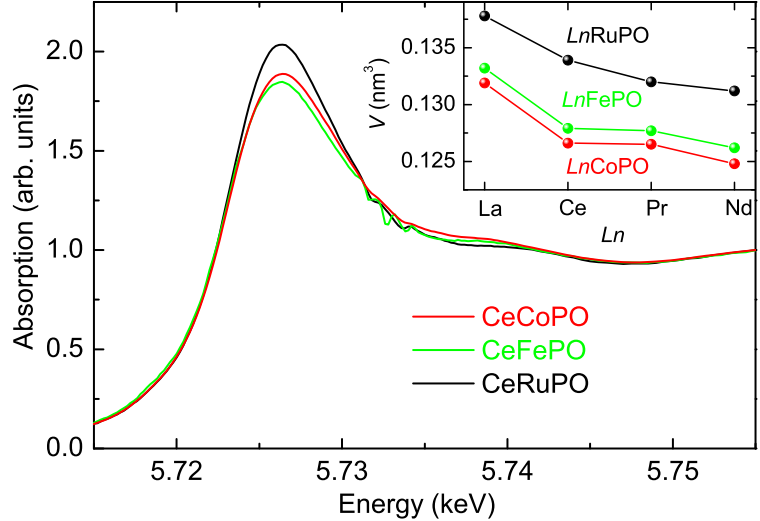}
 \caption{XANES spectra at the Ce $L_{\rm III}$ edge of CeTPO (T = Co, Fe, Ru) taken on powdered samples at $T=300$\,K. The data are normalized to 1 at 5.755\,keV. Inset: Cell volumes of the tetragonal phosphide oxides $LnT$PO \cite{Zimmer:1995}, \cite{Krellner}.}
\label{FigXAS}
\end{center}
\end{figure}

The Doniach picture can be used to classify the Ce$T$PO series with respect to the magnetism of the $4f$-electrons \cite{Doniach:1977}. From the inset of Fig.~\ref{FigXAS} it is evident that the volume of the unit cell decreases from $T$=Ru to Fe and further to Co. With decreasing $V$ the magnetism of the Ce-ions becomes more and more screened by the conduction electrons because of an increasing Kondo energy scale. Accordingly, we observe stable ferromagnetism of the Ce-ions in CeRuPO \cite{Krellner:2007} which has the largest $V_{\rm Ru}=0.1339$\,nm$^3$. CeFePO has a smaller $V_{\rm Fe}=0.1279$\,nm$^3$ and is close but already on the nonmangetic side of a FM instability \cite{Bruning:2008}. The volume related suppression of the magnetic order can also be observed in pressure experiments on single crystals of CeRuPO \cite{Macovei:2008}. 
Our conclusion of a larger screening of the $4f$ electrons as identified by a higher Kondo temperature $T_K\sim 40$\,K in CeCoPO is in nice continuation of this systematic trend since this compound has a further reduced volume $V_{\rm Co}=0.1266$\,nm$^3$ compared to CeFePO.

The superconductivity at $T_c=26$\,K in F-doped LaFeAsO came as a big surprise \cite{Kamihara:2008} and subsequently research on these systems reveals systems with $T_c$'s exceeding 50\,K in the elements with heavier $Ln$ \cite{Chen:2008}, \cite{ChenNature:2008}. The connection between a vanishing magnetic transition and the simultaneous formation of a superconducting state is reminiscent of the behavior in the cuprates and in the heavy Fermion systems, and therefore suggests the superconducting state in these doped RFeAsO systems to be of unconventional nature, too. In CeCoPO, we could not observe superconductivity down to 350\,mK, which is not so much surprising taking into account the FM Co-state. The magnetism might be easy to suppress either by doping or pressure, because an ordering temperature of $T_C=75$\,K is rather low for Co-ordering. Further research has to answer the question whether superconductivity appears in this compound after suppression of the FM order.

\section{Conclusion}
\label{Concl}
In summary, we have shown that CeCoPO presents ferromagnetic order of the Co-ions at $T_C=75$\,K. The $4f$-electrons are screened due to pronounced Kondo-interactions with a Kondo energy scale of order $T_K\sim 40$\,K revealed by the temperature dependence of the resistivity and an enhanced Sommerfeld-coefficient $\gamma\sim200$\,mJ/molK$^2$. In addition, we confirm the FM order in LaCoPO at $T_C=43$\,K.

\section*{Acknowledgements}
The authors thank P. Scheppan for energy dispersive x-ray analysis of the samples and N. Caroca-Canales and R. Weise for technical assistance in sample preparation. H. Rosner, E.~M.~Br\"uning, and M. Baenitz are acknowledged for valuable discussions.


\begin{thebibliography}{10}
\bibitem{Krellner}
We have redrawn the volumes of the unit cell from Ref.~\cite{Zimmer:1995};
  however, the volume plot there is different as Zimmer \textit{et al.} have
  used wrong absolute values for CeCoPO.
\end{thebibliography}
\end{document}